\documentclass[10pt]{article}

\usepackage[letterpaper, left=1in, top=1in, right=1in, bottom=1in, verbose, ignoremp]{geometry}

\usepackage{url}\RequirePackage[colorlinks,citecolor=blue, linkcolor=blue,urlcolor = blue]{hyperref}
\usepackage{latexsym,amssymb,amsmath,amsfonts,graphicx,color,fancyvrb,amsthm,enumerate,subfigure}
\usepackage[authoryear,round]{natbib}
\usepackage[dvipsnames]{xcolor}
\usepackage{xy}\xyoption{all} \xyoption{poly} \xyoption{knot}
\usepackage{float}
\usepackage{bm}
\usepackage{bbm}
\usepackage{multicol,multirow}
\usepackage{array}
\usepackage{relsize}
\usepackage{chngcntr}
\usepackage{etoolbox}

\usepackage{hyperref}

\def\eq#1{(\ref{#1})}

\def\beginmat{ \left( \begin{array} }
\def\endmat{ \end{array} \right) }

\def\cond{\, | \,}
\newcommand*\diff{\mathop{}\!\mathrm{d}}

\newcommand{\cH}{{\cal H}}
\newcommand{\cX}{{\cal X}}
\def\bR{{\mathbb R}}
\def\E{{\mathbb E}}
\newcommand{\T}{\intercal}

\makeatletter
\def\@biblabel#1{}
\makeatother

\makeatletter
\patchcmd{\NAT@citex}
  {\@citea\NAT@hyper@{%
     \NAT@nmfmt{\NAT@nm}%
     \hyper@natlinkbreak{\NAT@aysep\NAT@spacechar}{\@citeb\@extra@b@citeb}%
     \NAT@date}}
  {\@citea\NAT@nmfmt{\NAT@nm}%
   \NAT@aysep\NAT@spacechar\NAT@hyper@{\NAT@date}}{}{}

\patchcmd{\NAT@citex}
  {\@citea\NAT@hyper@{%
     \NAT@nmfmt{\NAT@nm}%
     \hyper@natlinkbreak{\NAT@spacechar\NAT@@open\if*#1*\else#1\NAT@spacechar\fi}%
       {\@citeb\@extra@b@citeb}%
     \NAT@date}}
  {\@citea\NAT@nmfmt{\NAT@nm}%
   \NAT@spacechar\NAT@@open\if*#1*\else#1\NAT@spacechar\fi\NAT@hyper@{\NAT@date}}
  {}{}

\makeatother

\begin{document}
\def\spacingset#1{\renewcommand{\baselinestretch}%
{#1}\small\normalsize} \spacingset{1}
\begin{flushleft}
{\Large{\textbf{Bayesian Approximate Kernel Regression with Variable Selection}}}
\newline
\\
Lorin Crawford\textsuperscript{1,$\dagger$},
Kris C. Wood\textsuperscript{2},
Xiang Zhou\textsuperscript{3,4,$\dagger$},
and Sayan Mukherjee\textsuperscript{1,5,6,7,$\dagger$}

\medskip

\bf{1} Department of Statistical Science, Duke University, Durham, NC, USA
\\
\bf{2} Department of Pharmacology \& Cancer Biology, Duke University, Durham, NC, USA
\\
\bf{3} Department of Biostatistics, University of Michigan, Ann Arbor, MI, USA
\\
\bf{4} Center for Statistical Genetics, University of Michigan, Ann Arbor, MI, USA
\\
\bf{5} Department of Computer Science, Duke University, Durham, NC, 27708
\\
\bf{6} Department of Mathematics, Duke University, Durham, NC, 27708
\\
\bf{7} Department of Bioinformatics \& Biostatistics, Duke University, Durham, NC, 27708
\\
\medskip
$\dagger$ Email: \url{lac55@stat.duke.edu}; \url{xzhousph@umich.edu}; \url{sayan@stat.duke.edu}
\end{flushleft}


\section*{Abstract}
Nonlinear kernel regression models are often used in statistics and machine learning because they are more accurate than linear models. Variable selection for kernel regression models is a challenge partly because, unlike the linear regression setting, there is no clear concept of an effect size for regression coefficients. In this paper, we propose a novel framework that provides an effect size analog for each explanatory variable in Bayesian kernel regression models when the kernel is shift-invariant --- for example, the Gaussian kernel. We use function analytic properties of shift-invariant reproducing kernel Hilbert spaces (RKHS) to define a linear vector space that: (i) captures nonlinear structure, and (ii) can be projected onto the original explanatory variables. This projection onto the original explanatory variables serves as an analog of effect sizes. The specific function analytic property we use is that shift-invariant kernel functions can be approximated via random Fourier bases. Based on the random Fourier expansion, we propose a computationally efficient class of Bayesian approximate kernel regression (BAKR) models for both nonlinear regression and binary classification for which one can compute an analog of effect sizes. We illustrate the utility of BAKR by examining two important problems in statistical genetics: genomic selection (i.e.~phenotypic prediction) and association mapping (i.e.~inference of significant variants or loci). State-of-the-art methods for genomic selection and association mapping are based on kernel regression and linear models, respectively. BAKR is the first method that is competitive in both settings.


\section{Introduction}\label{sec1}

In this paper, we formulate a nonlinear regression framework which simultaneously achieves the predictive accuracy of the most powerful nonlinear regression methods in machine learning and statistics, as well as provides an analog of effect sizes and probability of association for regression coefficients --- which are standard quantities in linear regression models.

Methodology and theory for variable selection is far more developed for linear regression models than nonlinear regression models. In linear models, regression coefficients (i.e.~the effect size of a covariate) provide useful information for variable selection. The magnitude and correlation structure of these effect sizes are used by various probabilistic models and algorithms to select relevant covariates associated with the response. Classic variable selection methods, such as forward and stepwise selection \citep{BromanSpeed}, use effect sizes to search for main interaction effects. Sparse regression models, both Bayesian \citep{ParkCasella} and frequentist \citep{Tibshirani:1996aa,LARS}, shrink small effect sizes to zero. Factor models use the covariance structure of the observed data to shrink effect sizes for variable selection \citep{West:2003aa,Hahn:2013aa}. Lastly, stochastic search variable selection (SSVS) uses Markov chain Monte Carlo (MCMC) procedures to search the space of all possible subsets of variables \citep{GeorgeMcCulloch}. All of these methods, except SSVS, use the magnitude and correlation structure of the regression coefficients explicitly in variable selection --- SSVS uses this information implicitly.

The main contribution of this paper is a (Bayesian) nonlinear regression methodology that is computationally efficient, predicts accurately, and allows for variable selection. The main technical challenge in formulating this novel method is defining and efficiently computing an analog of effect sizes for kernel regression models, a popular class of nonlinear regression models. Kernel regression models have a long history in statistics and applied mathematics \citep{Wahba:1990aa}, and more recently in machine learning \citep{Scholkopf:2002aa,Rasmussen}. There is also a large (partially overlapping) literature in Bayesian inference \citep{Wolpert:2007aa,MJ:2011aa,Chak:Ghos:Mall:2005}. 

The key idea we develop in this paper is that for shift-invariant kernels in the $p \gg n$ regime (i.e.~the number of variables $p$ is much larger than the number of observations $n$) there exists an accurate linear (in the covariates) approximation to the nonlinear kernel model that allows for the efficient computation of effect sizes. We specify a linear projection from the reproducing kernel Hilbert space (RKHS) to the space of the original covariates to implement the linear approximation. This linear transformation is based on the fact that shift-invariant kernels can be approximated by a linear expansion of random Fourier bases. The idea of random Fourier expansions was initially exploited to obtain kernel regression models with superior runtime properties in both training and testing \citep{Ben,Bazavan}. In this paper, we utilize the random Fourier bases to efficiently compute the analog of effect sizes for nonlinear kernel models.

The variable selection framework we develop is implemented as a Bayesian empirical factor model that scales to large datasets. Previous efforts to carry out variable selection in fully Bayesian kernel models have faced challenges when applied to large datasets due to either solving non-convex optimization problems \citep{Chapelle:2002aa,Rakotomamonjy,Rosasco} or due to sampling from Markov chains that typically mix poorly \citep{Chakraborty:2007aa,Chakraborty:2009aa}. Indeed, there have been recent works that attempt to overcome this problem with various approaches \citep{Snoek:2015aa,Gray-Davies:2016aa,Sharp:2016aa}. The main utility of our approach is that variable selection for nonlinear functions reduces to a factor model coupled with a linear projection. 

In section \ref{sec2}, we introduce properties of RKHS models and detail some of the basic functional analysis tools that allow for mapping from the RKHS of nonlinear functions to functions linear in the covariates. In section \ref{sec3}, we specify the Bayesian approximate kernel regression (BAKR) model for nonlinear regression with variable selection. Here, we also define the posterior probability of association analog (PPAA) which provides marginal evidence for the relevance of each variable. In section \ref{sec4}, we show the utility of our methodology on real and simulated data. Specifically, we focus on how our model addresses two important problems in statistical genetics: genomic selection and association mapping. Finally, we close with a discussion in section \ref{sec5}.


\section{Theoretical Overview}\label{sec2}

In this paper, we focus on nonlinear regression functions that belong to an infinite dimensional function space called a reproducing kernel Hilbert space (RKHS). The theory we develop in this section will help to formalize the following two observations in the $p \gg n$ setting: (i) the predictive accuracy of smooth nonlinear functions is typically greater than both linear functions and sharply varying nonlinear functions; (ii) in the high-dimensional setting, a smooth nonlinear function can be reasonably approximated by a linear function. In the remainder of this section, we develop a framework that we will use in section \ref{sec3} to define a linear projection from a RKHS onto the original  covariates. This projection will serve as an analog for effect sizes in linear models. Thorough reviews of the utility and theory of RKHS can be found in other selected works \citep[e.g.][]{Wolpert:2007aa,Bach:2017aa}.

\subsection{Reproducing Kernel Hilbert Spaces}\label{RKHS}

One can define an RKHS based on a positive definite kernel function, $k: {\mathcal X} \times {\mathcal X} \rightarrow \bR$, or based on the eigenfunctions $\{\psi_i\}_{i=1}^\infty$ and eigenvalues $\{\lambda_i\}_{i=1}^\infty$ of the integral operator defined by the kernel function, $\lambda_i \psi_i(\mathbf{u}) = \int_{\cX} k(\mathbf{u},\mathbf{v}) \psi_i(\mathbf{v}) \diff \mathbf{v}.$ For a Mercer kernel \citep{Merc:1909} the following expansion holds $k(\mathbf{u},\mathbf{v}) = \sum_{i=1}^\infty \lambda_i \psi_i(\mathbf{u}) \psi_i(\mathbf{v}),\label{mercer} $ and the RKHS can be alternatively defined as the closure of linear combinations of basis functions $\{\psi_i\}_{i=1}^\infty$,
\begin{align*}
\cH = \overline{\left\{ f \, |  \, f(\mathbf{x}) = \bm{\psi}(\mathbf{x})^{\T}\mathbf{c}, \, \forall \mathbf{x} \in \cX  \mbox{ and } \|f\|_\mathrm{K} < \infty \mbox{ with }  \|f\|_\mathrm{K}^2 = \sum_{i=1}^\infty \frac{\mathrm{c}_i^2}{\lambda_i^2} \right\}}.
\end{align*}
Here, $\|f\|_\mathrm{K}$ is the RKHS norm, ${\boldsymbol \psi}(\mathbf{x}) = \{\sqrt{\lambda}_i \psi_i(\mathbf{x})\}_{i=1}^\infty$ is a vector space spanned by the bases, and $\mathbf{c} = \{\mathrm{c}_i\}_{i=1}^\infty$ are the corresponding coefficients. The above specification of an RKHS looks very much like a linear regression model, except the bases are $\bm{\psi}(\mathbf{x})$ (rather than the unit basis), and the space can be infinite-dimensional. 

Kernel regression models in machine learning are often defined by the following penalized loss function \citep[Section 5.8]{Hastie01}
\begin{equation}
 \widehat{f} = \arg \min_{f \in \cH} \left[ \frac{1}{n} \sum_{i=1}^n L(f( \mathbf{x}_i),y_i) + \lambda \|f\|_\mathrm{K}^2 \right],\label{crit:regularization}
\end{equation}
where $\{(\mathbf{x}_i,\mathrm{y}_i)\}_{i=1}^n$ represents $n$ observations of covariates $\mathbf{x}_i \in \cX \subseteq \bR^p$ and responses $\mathrm{y}_i \in {\cal Y} \subseteq \bR$, $L$ is a loss function, and $\lambda > 0$ is a tuning parameter chosen to balance the trade-off between fitting errors and the smoothness of the function. The popularity of kernel models is that the minimizer of \eqref{crit:regularization} is a linear combination of kernel functions $k(\mathbf{u},\mathbf{v})$ centered at the observed data \citep{Scholkopf:2001aa}
\begin{equation}
 \widehat{f}(\mathbf{x}) = \sum_{i=1}^n \alpha_i k(\mathbf{x},\mathbf{x}_i),\label{representer:RKHS}
\end{equation}
where $\bm{\alpha} = \{\alpha_i \}_{i=1}^n$ are the corresponding kernel coefficients. The key point here is that the form of \eqref{representer:RKHS} turns an $\infty$-dimensional optimization problem into an optimization problem over $n$ parameters. We denote the subspace of the RKHS realized by the representer theorem as
\begin{align*} 
\cH_{_{\mathbf X}} =  \left \{f \, | \, f(\mathbf{x}) = \sum_{i=1}^n \alpha_i k(\mathbf{x},\mathbf{x}_i), \, \bm{\alpha} \in \bR^n \, \mbox{ and } \|f\|_\mathrm{K}^2 <  \infty \right\}.
\end{align*} 
We can also define the subspace $\cH_{_{\mathbf X}}$ in terms of the operator 
${\boldsymbol \Psi}_{_{\mathbf X}} =  [ {\boldsymbol \psi}(\mathbf{x}_1),\ldots,{\boldsymbol \psi}(\mathbf{x}_n)]$ with
\begin{align}
\cH_{_{\mathbf X}} = \left \{ f \, |  \, f(\mathbf{x}) = {\boldsymbol \Psi}_{_{\mathbf X}}^{\T}{\mathbf c} \, \mbox{ and } \|f\|_\mathrm{K}^2 <  \infty \right\}.\label{representer:Bases}
\end{align}
To extract an analog of effect sizes from our Bayesian kernel model we will use the equivalent representations \eqref{representer:RKHS} and \eq{representer:Bases}. Indeed, one can verify that $\mathbf{c} = {\boldsymbol \Psi}_{_{\mathbf X}}\boldsymbol{\alpha}$.

\subsection{Variable Selection in Kernel Models}

Variable selection in kernel models has often been formulated in terms of anisotropic functions
\begin{align*}
k_{\boldsymbol{\vartheta}}(\mathbf{u},\mathbf{v}) = k\Big((\mathbf{u} -\mathbf{v})^{\T}\mbox{Diag}(\boldsymbol{\vartheta})(\mathbf{u} - \mathbf{v})\Big), \quad \vartheta_j >0, \quad  j=1,\ldots,p
\end{align*}
where the vector $\boldsymbol{\vartheta}$ represents the weights each coordinate and is to be inferred from data. Optimization based approaches \citep{Chapelle:2002aa,Rakotomamonjy,Rosasco} implement variable selection by solving an optimization problem
\begin{align*}
\{\widehat{f}, \widehat{\boldsymbol{\vartheta}}\} = \arg \min_{f \in \cH_{\boldsymbol{\vartheta}}, \boldsymbol{\vartheta}} \left[ \frac{1}{n} \sum_{i=1}^n L(f( \mathbf{x}_i),\mathrm{y}_i) + \lambda \|f\|_{K_{\boldsymbol{\vartheta}}^2} \right],
\end{align*}
where $\cH_{\boldsymbol{\vartheta}}$ is the RKHS induced by the kernel $k_{\boldsymbol{\vartheta}}$, and the magnitude of $\widehat{\boldsymbol{\vartheta}}$ is evidence of the relevance of each variable. The joint optimization over $(\cH_{\boldsymbol{\vartheta}}, \boldsymbol{\vartheta})$ is a nonconvex problem and does not scale well with respect to the number of variables or the number of observations. In the case of Bayesian algorithms, the idea is to sample or stochastically search over the posterior distribution
\begin{align*}
p(\boldsymbol{\vartheta},\boldsymbol{\alpha} \mid \{\mathrm{y}_i,\mathbf{x}_i\}_{i=1}^n) \propto \exp\left\{- \sum_{i=1}^n L(f( \mathbf{x}_i),\mathrm{y}_i)\right\} \pi(\boldsymbol{\vartheta},\boldsymbol{\alpha}),
\end{align*}
where  $\pi(\boldsymbol{\vartheta},\boldsymbol{\alpha})$ is the prior distribution over the parameters and $\exp\left\{- \sum_{i=1}^n L(f( \mathbf{x}_i),\mathrm{y}_i)\right\}$ is the likelihood. Sampling over $\boldsymbol{\vartheta} \in \bR^p_+$ is challenging due to the complicated landscape, and Markov chains typically do not mix well in this setting \citep{Chakraborty:2007aa,Chakraborty:2009aa}. We will propose a very different approach to variable selection by projecting the RKHS onto linear functions with little loss of information. This projection operator will be based on random Fourier features.

\subsection{Random Fourier Features}

In this subsection, we specify an approximate kernel function that allows us to construct a projection operator between the RKHS and the original predictor space. This projection will allow us to define an analog to effect sizes. The key idea behind specifying the approximate kernel is the utilization of a previously developed randomized feature map \citep{Ben,Bazavan}. The approach we detail holds for kernel functions that are shift-invariant $k(\mathbf{u},\mathbf{v}) = k(\mathbf{u}-\mathbf{v})$ and integrate to one $\int k(\mathbf{z}) \diff \mathbf z = 1$ with $\mathbf{z} = \mathbf{u}-\mathbf{v}$. Bochner's theorem \citep{Bochner} states that this class of shift-invariant kernel functions satisfies the following Fourier expansion 
\begin{align}
k\left(\mathbf{x}_i-\mathbf{x}_j\right) = \int_{\mathbb{R}^p} f\left(\bm{\omega}\right)\exp\left\{\iota\,\bm{\omega}^{\T}\left(\mathbf{x}_i-\mathbf{x}_j\right)\right\}d\bm{\omega} = \mathbb{E}_{\bm{\omega}}\left[\eta_{\bm{\omega}}\left(\mathbf{x}_i\right)\eta_{\bm{\omega}}\left(\mathbf{x}_j\right)^*\right]\label{eq210},
\end{align} 
where $\eta_{\bm{\omega}}\left(\mathbf{x}_i\right) = \exp\left(\iota \, \bm{\omega}^{\T}\mathbf{x}_i\right)$, and the Fourier transform of the kernel function $f({\boldsymbol \omega})$ is a probability density
\begin{equation} \label{pdf}
f({\boldsymbol \omega}) = \int_{\cX} k({\mathbf x}) e^{-\iota 2 \pi {\boldsymbol \omega}^{\T}{\mathbf x}} \diff {\mathbf x}.
\end{equation}
More specifically, the eigenfunctions for these kernels can be thought of as Fourier bases.

Previous works have compared the performance of standard kernel regression models with an alternative Monte Carlo estimate representation using random bases \citep{Ben,Bazavan}
\begin{align*}
f(\mathbf{x}) = \sum_{i=1}^n \alpha_i k(\mathbf{x},\mathbf{x_i}), \quad f(\mathbf{x}) = \widetilde{\bm{\psi}}(\mathbf{x})^{\T}\mathbf{c},
\end{align*}
with $\widetilde{\bm{\psi}}(\mathbf{x})=[\widetilde{\psi}_1(\mathbf{x}),\ldots,\widetilde{\psi}_d(\mathbf{x})]^{\T}$ being a $d$-dimensional vector of Fourier bases with frequencies drawn from the corresponding density function in \eqref{pdf}. The specific Monte Carlo approximation is formulated as follows
\begin{align}
\label{randexpan}
&\bm{\omega}_\ell \stackrel{iid}{\sim} f({\boldsymbol \omega}), \quad \mathrm{b}_\ell \stackrel{iid}{\sim} U[0,2\pi], \quad  \ell=1,\ldots,d &\nonumber \\
&\bm{\Omega} = \left[\bm{\omega}_1,\ldots,\bm{\omega}_d\right] \in \mathbb{R}^{p\times d}, \quad \mathbf{b} = \left[\mathrm{b}_1,\ldots,\mathrm{b}_d\right] \in \mathbb{R}^d, &\\
&\widetilde{\bm{\psi}}(\mathbf{x}_i)^{\T} =  \sqrt{\frac{2}{d}} \cos \left(\mathbf{x}_i\bm{\Omega} + \mathbf{b}\right), & \nonumber
\end{align}
where $\cos(\bm{v})$ denotes an element-wise cosine transformation, and (in the case of the Gaussian kernel) $f({\boldsymbol \omega})$ is a $p$-dimensional multivariate normal. Note that the adaptation of \eq{randexpan} to accommodate other shift-invariant kernel functions is straightforward and only requires sampling from a different $f({\boldsymbol \omega})$. We may now consider a Monte Carlo estimate of the kernel function where
\begin{align}
k\left(\mathbf{x}_i-\mathbf{x}_j\right) = \bm{\psi}\left(\mathbf{x}_i\right)^{\T}\bm{\psi}\left(\mathbf{x}_j\right) \approx \widetilde{\bm{\psi}}\left(\mathbf{x}_i\right)^{\T}\widetilde{\bm{\psi}}\left(\mathbf{x}_j\right) = \widetilde{k}\left(\mathbf{x}_i-\mathbf{x}_j\right)\label{eq29}.
\end{align}
We can also specify a matrix $\widetilde{\bm{\Psi}} = [\widetilde{\bm{\psi}}(\mathbf{x}_1),\ldots,\widetilde{\bm{\psi}}(\mathbf{x}_n)]$ with a corresponding approximate kernel matrix $\widetilde{\mathbf{K}} = \widetilde{\bm{\Psi}}^{\T}\widetilde{\bm{\Psi}}$. It has been previously shown, using the strong law of large numbers, that this approximation converges to the exact kernel almost surely as the random sampling size $d$ goes to infinity \citep{Ben}. In this paper, we  set $d = p$ for two reasons. First, since our focus is in the $p\gg n$ setting, the number of terms in the Fourier expansion will be large and the resulting difference between the actual smooth nonlinear kernel function and its Monte Carlo approximation will be small \citep{Bach:2017aa}. Second, setting $d=p$ allows us to define a linear map from the $n$-dimensional space of kernel functions to a $p$-dimensional projection onto the original explanatory variables.

Note that in the construction we have proposed, the approximate kernel function is conditional on the random quantities $\{\bm{\omega}_\ell, \mathrm{b}_\ell\}_{\ell=1}^d$. This can be made explicit by the notation $\widetilde{k}_{\bm{\omega}, \mathrm{b}} \equiv \widetilde{k}(\mathbf{u},\mathbf{v}) \mid \{\bm{\omega}_\ell, \mathrm{b}_\ell \}_{\ell=1}^d$. Thus, \eqref{randexpan} can be thought of as a prior specification over $\{\bm{\omega}_\ell, \mathrm{b}_\ell\}_{\ell=1}^d$. In a fully Bayesian model, we can also infer posterior summaries on these random kernel parameters. In this paper, we will fix the Fourier bases and avoid sampling the random parameters to minimize computational burden as estimating the joint posterior distribution of $\{\bm{\omega}_\ell, \mathrm{b}_\ell\}_{\ell=1}^d$ will significantly increase computational cost. In addition, we know that the Monte Carlo error between the actual kernel and approximate kernel is very small \citep{Ben,Bazavan,Bach:2017aa}. Therefore, the effect of posterior estimates for $\{\bm{\omega}_\ell, \mathrm{b}_\ell \}_{\ell=1}^d$ on predictive accuracy and inference will be minimal. For the rest of the paper, we consider the following approximation when we specify the approximate kernel: $\widetilde{k}(\mathbf{u},\mathbf{v}) \equiv \widetilde{k}_{\bm{\omega}, \mathrm{b}}(\mathbf{u},\mathbf{v})$. We will also assume that for two runs of the model the difference between approximate kernels will be negligible. We provide empirical evidence of this assertion via a simulation study in section \ref{sec4}.  


\section{Bayesian Approximate Kernel Regression}\label{sec3}

We now state the general framework for Bayesian approximate kernel regression (BAKR) using random Fourier features. Extensions for binary classification and mixed effect modeling can be found in Supporting Information. We will also specify prior distributions that induce accurate and interpretable estimation of effect sizes in the original covariate space.

\subsection{Generalized Approximate Kernel Models}

The loss function in the penalized estimator in equation \eqref{crit:regularization} corresponds to a negative conditional log-likelihood for many penalized estimators \citep[e.g.][]{Mallick:2005aa}. For a standard linear model the following holds 
\begin{align}
\mathbf{y} \sim p(\mathbf{y}\cond\bm{\mu}) \quad \text{with} \quad \bm{\mu} = \mathbf{X}\bm{\beta}.\label{eq31}
\end{align}
Similarly, we can specify a generalized kernel model (GKM) \citep{MJ:2011aa} based on the random kernel expansion in \eqref{randexpan}. Technically this is a conditional model as both $\widetilde{\mathbf{K}}$ and $\widetilde{\bm{\Psi}}$ are conditional on $\{\bm{\omega}_\ell, \mathrm{b}_\ell\}_{\ell=1}^d$. In addition, we also only consider one realization of $\{\bm{\omega}_\ell, \mathrm{b}_\ell\}_{\ell=1}^d$, as noted in the previous section. Again we assume that, for two runs of the model, the difference between approximate kernel matrices will be negligible. The GKM framework takes on the following form
\begin{align}
\mathbf{y} \sim p(\mathbf{y}\cond\bm{\mu}) \quad \text{with} \quad g^{-1}(\bm{\mu}) = \widetilde{\bm{\Psi}}\mathbf{c},\label{eq32}
\end{align}
where $g$ is a link function. The above can also be written in terms of the approximate kernel matrix
\begin{align}
\mathbf{y} \sim p(\mathbf{y}\cond\bm{\mu}) \quad \text{with} \quad g^{-1}(\bm{\mu}) = \widetilde{\mathbf{K}}\bm{\alpha},\label{eq33}
\end{align}
where $\bm{\alpha} = \widetilde{\mathbf{K}}^{-1}\widetilde{\bm{\Psi}}^{\T}\mathbf{c}$. We will refer to the model specified in \eqref{eq33} as a \textit{generalized approximate kernel model}. The generalized models provide a unifying framework for kernel-based regression and classification \citep[e.g.][]{Mallick:2005aa,Chakraborty:2007aa,Chakraborty:2009aa,Chak:Ghos:Mall:2005}. Depending on the application of interest, one may specify an appropriate likelihood and link function. For instance, in the regression case, the likelihood is specified as a normal distribution and the link function is the identity. In this paper, we will use this framework for regression and classification applications. 
 
\paragraph{Approximate Kernel Factor Models.} Since the kernel matrix $\mathbf{K}$ is symmetric and positive (semi) definite, one can apply a variety of factor model methodologies \citep[e.g.][]{West:2003aa,Campos:2010aa}. The approximate kernel matrix $\widetilde{\mathbf{K}}$ is also symmetric and positive (semi) definite, and thus satisfies the following spectral decomposition $\widetilde{\mathbf{K}} = \widetilde{\mathbf{U}}\widetilde{\bm{\Lambda}}\widetilde{\mathbf{U}}^{\T},$ where $\widetilde{\mathbf{U}}$ is an $n\times n$ orthogonal matrix of eigenvectors and $\widetilde{\bm{\Lambda}} = \text{Diag}(\lambda_1,\ldots,\lambda_n)$ is a diagonal matrix of eigenvalues sorted in decreasing order. For numerical stability and reduction of computational complexity, eigenvectors corresponding to small eigenvalues can be truncated. Therefore, without loss of generality we can consider $\widetilde{\mathbf{U}}$ as a $n\times q$ matrix of eigenvectors and $\widetilde{\bm{\Lambda}}$ as a $q\times q$ diagonal matrix of the top $q$ eigenvalues. With this factor model representation, we rewrite \eq{eq33} as 
\begin{align}
\mathbf{y} \sim p(\mathbf{y}\cond\bm{\mu}) \quad \text{with} \quad g^{-1}(\bm{\mu}) = \widetilde{\mathbf{U}}\bm{\theta},\label{eq34}
\end{align}
where $\bm{\theta} = \widetilde{\bm{\Lambda}}\widetilde{\mathbf{U}}^{\T}\bm{\alpha}$ is a $q$-dimensional vector of latent regression parameters or coefficients. The reduced orthogonal factor matrix is an orthonormal representation of the nonlinear relationship between samples and allows for further dimension reduction from $n$ to $q$ parameters. This representation can greatly speed up estimation of model parameters, especially when $n$ is large in the Bayesian paradigm. In this work, we will choose $q$ to correspond to the eigenvalues that explain a majority of the cumulative variance in $\widetilde{\mathbf{K}}$ --- noting that more variance explained typically equates to better model performance. Later, we explore an empirical sensitivity analysis for this choice of $q$ via a simulation study. Other practical suggestions for how to choose $q$ can be found in previous works \citep{West:2003aa,Campos:2010aa}.

\subsection{Projection onto Explanatory Variables and Effect Size Analog}

We now define the analog to effect sizes for nonlinear kernel models. We first briefly outline effect sizes for linear models. In linear models, a natural interpretation of the effect size of a coefficient is the magnitude of the projection of the design matrix $\mathbf{X}$ onto the expectation of the response vector $\E [\mathbf y]$: ${\bm{\beta}} = \mbox{Proj}({\mathbf X},\E[\mathbf{y}])$. In practice, one does not have access to $\E [\mathbf{y}]$; hence, the projection will be specified as $\mbox{Proj}({\mathbf X},\mathbf{y})$, with the choice of loss function/noise model as well as priors or regularizations specifying the exact form of the projection. The standard projection operation is $\mbox{Proj}(\mathbf{X},\mathbf{y}) = \mathbf{X}^{\dagger} {\mathbf y}$, where $\mathbf{X}^{\dagger}$ is the Moore-Penrose generalized inverse which, in the case of a full rank design matrix, is $\mathbf{X}^{\dagger} = (\mathbf{X}^{\T} \mathbf{X})^{-1} \mathbf{X}^{\T}$ and leads to the standard least-squares regression coefficient estimates. For Bayesian procedures, priors over the parameters $\bm{\beta}$ induce a distribution on the projection procedure $\mbox{Proj}(\mathbf{X}, \mathbf{y})$. 

Our definition for the effect size analog is based on the same idea of projecting a nonlinear function onto the design matrix. Specifically, consider a nonlinear function evaluated on the observations $\E[\mathbf{y}] = \mathbf{f} = [f(\mathbf{x}_1) \cdots f(\mathbf{x}_n)]^\T$. We define the \textit{effect size analog} as determined by the coefficients that result from projecting the design matrix $\mathbf{X}$ onto the nonlinear response vector $\mathbf{f}$,  
\begin{align}
\widetilde{\bm{\beta}} = \mbox{Proj}({\mathbf X},{\mathbf f})\label{geff}.
\end{align}
The main purpose of the random Fourier bases and the approximate kernel factor model is to specify a projection operation that is well defined, efficient, and results in robust estimates. The remainder of this subsection consists of defining this projection operation and its practical calculation of which requires three sets of coefficients: (1) the coefficients $\mathbf{c}$ of the random Fourier bases;  (2) the coefficients $\bm{\theta}$ from the empirical factor representation of the probabilistic model; and (3) the coefficients that determine the effect size analog $\widetilde{\bm{\beta}}$. Recall that $\bm{\alpha} = \widetilde{\mathbf{K}}^{-1}\widetilde{\bm{\Psi}}^{\T}\mathbf{c}$ and $\bm{\theta} = \widetilde{\bm{\Lambda}}\widetilde{\mathbf{U}}^{\T}\bm{\alpha}$. We use the transformations defined in equations \eq{eq32}-\eq{eq34} to specify the relationship between $\bm{\theta}$  and $\mathbf{c}$ as the following: $\mathbf{c} = (\widetilde{\bm{\Lambda}} \widetilde{\mathbf{U}}^{\T}\widetilde{\mathbf{K}}^{-1} \widetilde{\bm{\Psi}}^{\T})^{-1} \bm{\theta}.$ Following the formulation of the effect size analog in equation \eqref{geff}, we specify the projection of $\mathbf{f} = \widetilde{\bm{\Psi}}^{\T} \mathbf{c}$ onto the design matrix as the linear map
\begin{align}
\widetilde{\bm{\beta}} = \mathbf{X}^{\dagger}\widetilde{\bm{\Psi}}^{\T} \mathbf{c},\label{eq36}
\end{align}
where the above follows directly from equations \eqref{eq32} and \eqref{geff}. The argument for why $\widetilde{\bm{\beta}}$ is an effect size analog for the kernel regression model is that, on the $n$ observations, $\mathbf{K}\bm{\alpha} \approx \mathbf{X}\widetilde{\bm{\beta}}$. It should be clear that a variety projection procedures can be specified corresponding to various priors and loss functions, and a systematic study elucidating which projections are efficient and robust is of great interest. We provide strong evidence in this paper that the projection procedure specified by equation \eqref{eq36} results in efficient and robust inference. 

Throughout the rest of the paper, all of our prior specifications and modeling efforts will be placed on the empirical factor kernel coefficients $\bm{\theta}$. Ideally, we would like a one-to-one map between the $q$-dimensional factor regression coefficients and the $p$-dimensional effect size analogs. In Supporting Information, we formally show that the map from $\mathbf{c}$ to $\widetilde{\bm{\beta}}$ is injective modulo the null space of the design matrix $\mathbf{X}$. This result is unsurprising since, in the classical linear regression case, two different coefficient vectors will result in the same response vector if the difference between the vectors is in the null space of $\mathbf{X}$.

\subsection{Bayesian Hierarchical Model Specification}

We first restate the approximate kernel model for regression, 
\begin{align}
\mathbf{y} = \widetilde{\mathbf{K}}\bm{\alpha} + \bm{\varepsilon}, \quad \bm{\varepsilon} \sim \text{MVN}(\bm{0},\tau^2\mathbf{I}).\label{eq41}
\end{align}
We will implicitly specify priors over the coefficients $\bm{\alpha}$. In general, we would like these priors to adaptively shrink based on sample size. A natural way to induce sample size dependent shrinkage and account for covariance structure is using a g-prior. We adopt a standard approach in kernel based Bayesian models that uses this class of priors to shrink the parameters proportional to the variance of the different principal component directions on the original covariate design space \citep[e.g.][]{Campos:2010aa}. We now give a hierarchical specification for the nonlinear regression model in \eq{eq41} in terms of its empirical factor representation, as derived in \eq{eq34}. Namely,
\begin{equation}
\begin{aligned}
\mathbf{y} &= \widetilde{\mathbf{U}}\bm{\theta}+\bm{\varepsilon}, \quad \bm{\varepsilon} \sim \text{MVN}(\bm{0},\tau^2\mathbf{I}),\\
\bm{\theta} &\sim \text{MVN}(\boldsymbol{0},\sigma^2\widetilde{\mathbf{\Lambda}}),\label{eq42}\\
\sigma^2,\tau^2 &\sim \mbox{Scale-inv-}\chi^2(\nu,\phi).
\end{aligned}
\end{equation}
For the residual variance parameter $\tau^2$ we specify a scaled-inverse chi-square distribution with degrees of freedom $\nu$ and scale $\phi$ as hyper-parameters. The regression parameters are assigned a multivariate normal prior. The idea of specifying prior distributions in the orthogonal space on $\bm{\theta}$, instead of $\boldsymbol{\alpha}$, is referred to as the Silverman g-prior \citep{MJ:2011aa}. The parameter $\sigma^2$ is a shrinkage parameter and is also assumed to come from a scaled-inverse chi square distribution. The joint specification for $\sigma^2$ and $\bm{\theta}$ has the advantage of being invariant under scale transformations and induces a heavier tailed prior distribution for $\bm{\theta}$ when marginalizing over $\sigma^2$. In particular, $\sigma^2$ allows for varying the amount of shrinkage in each of the orthogonal factors of $\widetilde{\mathbf{K}}$. This mitigates the concern in principal component regression where the dominant factors may not be most relevant to the modeling problem.

\paragraph{Posterior Inference and Sampling.} Given the model specification in \eq{eq42}, we use a Gibbs sampler to draw from the joint posterior  $p(\bm{\theta},\sigma^2,\tau^2\cond \mathbf{y})$. In many applications, the kernel function is indexed by a bandwidth or smoothing parameter $h$, with $k_h(\mathbf{u},\mathbf{v})$ \citep[e.g.][]{MJ:2011aa, Chak:Ghos:Mall:2005}. For example, the Gaussian kernel can be specified as $k_h(\mathbf{x}_i,\mathbf{x}_j) = \exp\{-h\|\mathbf{x}_i-\mathbf{x}_j\|^2\}$. This bandwidth parameter can be inferred; however, posterior inference over $h$ is slow, complicated, and mixes poorly. We work with a fixed bandwidth, and hence a fixed approximate kernel function, allowing us to avoid this computational cost \citep{Ben,Bazavan}. The Gibbs sampler we propose consists of the following closed form conditional densities:
\begin{enumerate}
\item[(1)] $\bm{\theta} \mid \sigma^2,\tau^2, \mathbf{y}  \sim \text{MVN}(\mathbf{m}^{*},\mathbf{V}^{*})$ with $\mathbf{V}^{*} = \tau^2\sigma^2(\tau^2\widetilde{\bm{\Lambda}}^{-1}+\sigma^2\mathbf{I}_q)^{-1}$ and $\mathbf{m}^* = \tau^{-2}\mathbf{V}^{*}\widetilde{\mathbf{U}}^{\T}\mathbf{y}$;
\item[(2)] $\widetilde{\bm{\beta}} = \mathbf{X}^{\dagger}\widetilde{\bm{\Psi}}^{\T}(\widetilde{\bm{\Lambda}}\widetilde{\mathbf{U}}^{\T}\widetilde{\mathbf{K}}^{-1}\widetilde{\bm{\Psi}}^{\T})^{-1}\bm{\theta}$;
\item[(3)] $\sigma^2 \mid \bm{\theta},  \tau^2, \mathbf{y}  \sim \mbox{Scale-inv}-\chi^2(\nu^*_{\sigma},\phi^{*}_{\sigma})$ where $\nu^*_{\sigma} = \nu + q$ and $\phi^{*}_{\sigma} = \nu^{*-1}_{\sigma}(\nu\phi+\bm{\theta}^{\T}\widetilde{\bm{\Lambda}}^{-1}\bm{\theta})$;
\item[(4)] $\tau^2 \mid  \bm{\theta},  \sigma^2, \mathbf{y} \sim \mbox{Scale-inv}-\chi^2
(\nu^*_{\tau},\phi^{*}_{\tau})$ where $\nu^*_{\tau} = \nu+n$ and $\phi^{*}_{\tau} = \nu^{*-1}_{\tau}(\nu\phi+\mathbf{e}^{\T}\mathbf{e})$, with $\mathbf{e} = \mathbf{y} - \widetilde{\mathbf{U}}\bm{\theta}$.
\end{enumerate}
Note that the second step is deterministic and maps back to the effect size analogs. Iterating the above procedure $T$ times results in the following set of posterior draws: $\{\bm{\theta}^{(t)},\sigma^{2(t)},\tau^{2(t)}, \widetilde{\bm{\beta}}^{(t)}\}_{t=1}^T$. 

In many applications, it is of key interest to select a sparse set of variables based on the posterior samples of the model effect sizes. Since BAKR utilizes an induced prior on $\widetilde{\bm{\beta}}$ that is non-sparse, we cannot directly use standard quantities such as the posterior inclusion probability (PIP)  or the posterior probability of association (PPA) \citep{Stephens_Nature} to select variables with non-zero effects. Instead, we advertise the use of hard thresholding and define a quantity termed the ``posterior probability of association analog'' (PPAA) to perform variable selection. The PPAA is effectively an analog to the PPA and the PIP. More specifically, it uses hard thresholding which is both computationally and statistically efficient in high dimensional settings, and correctly controls false discovery rates for reasonably sized datasets (see Supporting Information for details). Briefly, for a given threshold $z_{j^*}$, the PPAA may be represented as 
\begin{align}
\widetilde{\gamma}_j = \begin{cases}
1 \quad &\mbox{if} \quad |\widetilde{\beta}_j|\ge z_{j^*}\\
0 \quad &\mbox{if} \quad \text{Otherwise}
\end{cases}
\quad \mbox{for} \quad j = 1,\ldots, p\label{HT}
\end{align}
where $\widetilde{\gamma}_j$ effectively represents an indicator that predictor variable $j$ is associated with the response. In statistical genetics applications, we can define candidate causal variables as those covariates that satisfy $\left\{\widetilde{\gamma}: \mathrm{Pr}[\widetilde{\gamma} = 1\mid \mathbf{y}] > r\right\}$. In practice, $r$ may be chosen subjectively \citep[e.g.][]{Hoti}, or taken to be $r$ = 0.5 in order to obtain an equivalence of a Bayesian ``median probability model''. Another option is to define $r$ through k-fold permutation to find an effective predictor-wide threshold. For any set of significant variables, further analyses may be carried out involving the relative costs of false positives and false negatives to make an explicitly reasoned decision about which predictors to pursue \citep{Stephens_Nature}. In Supporting Information, we show how PPAAs can offer theoretical guarantees of a sparse solution under certain regularity conditions. We also provide empirical evidence that validates our approach. 

\paragraph{Prediction.} Prediction using BAKR is very similar to prediction in other Bayesian parametric models. When an $n^*\times p$ dimensional out of sample test set $\mathbf{X}^*$ is observed, a quantity of interest is the posterior predictive distribution of the vector $\mathbf{y}^*$. Given the posterior draws of the model's parameters, samples are drawn from the posterior predictive distribution as $\{\mathbf{y}^{*(t)} = \mathbf{X}^*\widetilde{\bm{\beta}}^{(t)}\}_{t=1}^T.$ With sampled parameters at each iterate, we can generate posterior predictive quantities and Monte Carlo approximations of marginal predictive means across a range of new sample values.

Note that there are a few differences between the posterior predictive intervals computed in standard Bayesian nonparametric models (e.g.~Gaussian process regression), and the intervals computed by BAKR. In the case of standard Bayesian nonparametric models, the predictive distribution is a normal distribution that depends on a kernel matrix that is evaluated at all $n+n^*$ observations in a combined design matrix (i.e.~$\mathbf{X} \cup \mathbf{X}^*$). Thus, whenever new samples are observed, the model is respecified and the posterior must to be recomputed \citep[e.g.][]{Mallick:2005aa}. However, in BAKR, the implied posterior distribution on $\widetilde{\bm{\beta}}$ depends only on the kernel matrix evaluated at the training data. This serves as an advantage since, when new samples are observed, we can obtain the posterior predictive distribution using the estimates of the effect size analogs.

\subsection{Capturing Interaction Effects}\label{sec44}

A main motivation for  BAKR is variable selection for nonlinear functions. In genetics applications, understanding how nonlinear interactions between genes influence the genetic architecture of traits and variation in phenotypes is of great interest \citep{Mackay:2014aa}. These nonlinear or non-additive interactions are commonly referred to as ``epistasis". One approach to modeling epistatic effects is to explicitly include interactions between all combinations of covariates in the model, and then obtain posterior samples (or point estimates) of the interactions. However, \textit{a priori}, it is almost impossible to know how many higher order interaction terms are needed to fully capture the variation in an observed response. A more serious practical issue is that combinatorial growth in the number of interactions results in a parametric linear model that is computationally infeasible to run. Moreover, variance in the estimates of the regression parameters explodes as the model space grows.

It has been shown that the Taylor series expansion of the Gaussian kernel function enumerates all higher order interaction terms between covariates \citep{Jiang:2015aa}, thus alleviating potential combinatorial and computational concerns. The approximate Gaussian kernel shares this same property. Note that we can write the Gaussian kernel as a product of three terms
\begin{align*}
k(\mathbf{u},\mathbf{v}) = \exp\{-h\|\mathbf{u}-\mathbf{v}\|^2\} = \exp\{-h\|\mathbf{u}\|^2\}\exp\{-h\|\mathbf{v}\|^2\}\exp\{-h\langle\mathbf{u},\mathbf{v}\rangle\}.
\end{align*}
The Taylor series expansion of the third term makes it clear why the Gaussian kernel can be thought of as a collection of $m$\textsuperscript{th} higher order interaction terms between variables
\begin{align*}
\exp\{-h\langle\mathbf{u},\mathbf{v}\rangle\} = \sum_{m=0}^\infty \frac{h^m}{m!}(\mathbf{u}^{\T}\mathbf{v})^m \mbox{ where } (\mathbf{u}^{\T}\mathbf{v})^m =  \sum_{j \in [p]^m} \left(\prod_{i=1}^{m} \mathbf{u}_{j_i}\right) \left(\prod_{i=1}^{m} \mathbf{v}_{j_i}\right).
\end{align*}
Here, $[p]^m$ is the set of $m$ coordinate subsets of the $p$ coordinates \citep{Cotter:2011aa}.


\section{Results}\label{sec4}

We illustrate the utility of BAKR on both simulated and real data. The motivation for both sets of examples is to understand the performance of BAKR on two important problems in statistical genetics: genomic selection and association mapping. First, we use simulations to assess the Monte Carlo error between the actual kernel matrix and the approximate kernel matrix. Specifically, we investigate how much of the performance variation between runs of BAKR is due to posterior estimation via the Gibbs sampler, versus how much is due to the sampling of the approximate kernel function. Next, we assess the modeling performance of BAKR by using two simulation scenarios corresponding to two different types of genetic architectures. Here, the goal is to show that BAKR performs association mapping as well as the most commonly used Bayesian variable selection methods, and predicts phenotypes as accurately as the best nonparametric models. Finally, we will assess BAKR's prediction and association mapping abilities in two real datasets. The first is a stock mouse dataset from the Wellcome Trust Centre for Human Genetics, and the second is a large human dataset from the Wellcome Trust Case Control Consortium (WTCCC).

\subsection{Simulations: Approximation Error Assessment}

In this subsection, we use simulations to assess the Monte Carlo error between the actual kernel matrix and the approximate kernel matrix. Specifically, we use a simulated matrix $\mathbf{X}$ with $p = 2000$ covariates to create continuous outcomes using the following generating polynomial model: $\mathbf{y} = \mathbf{X}^3 \mathbf{b}+\bm{\varepsilon}$ where $\bm{\varepsilon}\sim\text{MVN}(\mathbf{0},\mathbf{I})$ and $\mathbf{X}^3 = \mathbf{X}\circ\mathbf{X}\circ\mathbf{X}$ is the element-wise third power of $\mathbf{X}$. We assume that the first 100 covariates are relevant to the response with $\mathbf{b}_{1:100}\sim\text{MVN}(\mathbf{0},\mathbf{I})$, while the remainder are assumed to have zero effect. Here, we consider sample sizes $n = \{500, 750, 1000\}$, where we analyzed 100 different datasets in each case. Within each individual dataset, we run BAKR using an approximate Gaussian kernel matrix 100 different times in order to get a clear illustration of the variation in performance between runs of the model. We also implement BAKR using an exact Gaussian kernel 100 different times as a direct comparison. The idea here is that we want to investigate how much of the variation between runs of BAKR is due to posterior estimation via the Gibbs sampler, versus how much is due to the sampling of the approximate kernel function. In order to do this, within each iterative run, we compute a predicted value $\widehat{\mathbf{y}}$ for both models and then calculate their respective $\text{R}^2$. We will treat the variance of the computed $\text{R}^2$ across runs as a quantity to measure error. In Table S1, we decompose the variance of $\text{R}^2$ into the proportion due to MCMC and the proportion due to the approximate kernel (see Supporting Information). For each sample size considered, we see that the average variance between runs of BAKR is negligible (i.e.~$7.02\times10^{-6}$, $2.26\times10^{-6}$, and $1.04\times10^{-6}$ for sample sizes $n =$ 500, 750, and 1000, respectively). We also see that the proportion of this variance due to sampling the approximate kernel function is also small. 

\subsection{Simulations: Model Assessment}

To assess the power of BAKR, we consider simulation designs similar to those proposed by previous genetic analysis studies \citep[e.g.][]{Wan:2010aa}. First, we assume that genetic effects explain 60\% of the total variance in the response --- this is analogous to assuming that the broad-sense heritability of a trait is known (i.e.~$\mathrm{H}^2=0.6$) \citep{Zhou:2013aa}. Next, we use a simulated genotype matrix $\mathbf{X}$ with $n = 500$ samples and $p = 2000$ single-nucleotide polymorphisms (SNPs) to create continuous phenotypes that mirror genetic architectures affected by a combination of linear (additive) and interaction (epistatic) effects. Specifically, we randomly choose 50 causal SNPs that we classify into two distinct groups: (i) a set of 25 additive SNPs, and (ii) a set of 25 interaction SNPs. 

The additive effect sizes of the group 1 causal SNPs come from a standard normal distribution or $\textbf{b}\sim\mbox{MVN}(\bm{0},\mathbf{I})$. Next, we create a separate matrix $\mathbf{W}$ which holds all pairwise interactions between the group 2 causal SNPs. The corresponding interaction effect sizes are also drawn as $\textbf{a}\sim\mbox{MVN}(\bm{0},\mathbf{I})$. We scale both the additive and interaction genetic effects so that collectively they explain a fixed proportion of broad-sense heritability. Namely, the additive effects make up $\rho$\%, while the pairwise interactions make up the remaining $(1-\rho)$\%. Once we obtain the final effect sizes for all causal SNPs, we draw normally distributed random errors as $\bm{\varepsilon}\sim\mbox{MVN}(\bm{0},\mathbf{I})$ to achieve the target H\textsuperscript{2}. Finally, phenotypes are then created by summing over all effects using the simulation model: $\mathbf{y} = \mathbf{X}\mathbf{b}+\mathbf{W}\mathbf{a} +\bm{\varepsilon}$.  

We consider two simulation scenarios that depend on the parameter $\rho$, which measures the proportion of H\textsuperscript{2} that is contributed by the effects of the first and second groups of causal SNPs. Specifically, the phenotypic variance explained (PVE) by the additive genetic effects is said to be $\mbox{V}(\mathbf{X}\mathbf{b})= \rho\mbox{H}^2$, while the PVE of the pairwise interaction genetic effects is given as $\mbox{V}(\mathbf{W}\mathbf{a})=(1-\rho)\mbox{H}^2$. Here, we choose the set $\rho = \{0.2,0.8\}$ corresponding to scenarios I and II, respectively. In the particular case where $\rho = 0.2$, the epistatic effects are assumed to dominate the broad-sense heritability of the simulated phenotypes. The alternative case in which $\rho = 0.8$ is a scenario where the PVE of the simulated responses are dominated by additive effects. We analyze 100 different simulated datasets for each value of $\rho$. 

\paragraph{Variable Selection.} Here, we are interested in examining the power of BAKR to effectively identify the additive and interacting causal SNPs under different genetic architectures. We compare our proposed method to four other standard Bayesian variable selection models, which we implement using the R package BGLR \citep{Perez:2014aa}. More specifically, these methods includ: (a) Bayesian Ridge Regression
\citep{Hoerl:2000aa} with a fixed prior variance for each variable; (b) Bayesian Lasso \citep{ParkCasella}; (c) Bayesian linear mixed model (LMM) with the random effects specified as a multivariate normal distribution with mean vector $\bm{0}$ and covariance matrix $\mathbf{K} = \mathbf{X}\mathbf{X}^{\T}/p$, where $\mathbf{K}$ here is referred to as a linear or additive kernel matrix \citep{Jiang:2015aa}; (d) a common spike and slab prior model, also commonly known as the Bayes C$\pi$ model or Bayesian variable selection model \citep{Habier:2011aa}, which is specified as a mixture of a point mass at zero and a diffuse normal centered around zero. 

In order to illustrate the robustness of BAKR, we apply our method while approximating Gaussian kernels with bandwidth parameter values $h=\{5,2,1,0.5,0.01\}$, where each value of $h$ represents a varying degree of kernel density smoothness. The goal is to show that the power of BAKR is robust under reasonable choices for $h$. In each case, we assume that BAKR utilizes its empirical kernel factor formulation while including all eigenvectors explaining 95\% of the cumulative variance in the top $q$ eigenvalues, and set the model hyper-parameters to $\nu = 5$ and $\phi = 2/5$. In Supporting Information, we conduct an empirical analysis where we assess the sensitivity of BAKR to the choice of $q$ (see Figure S1).

\begin{figure}[H]
\centering
\subfigure[Scenario I ($\rho=0.2$): Group 1 SNPs]{
   \includegraphics[width = 0.4\textwidth]{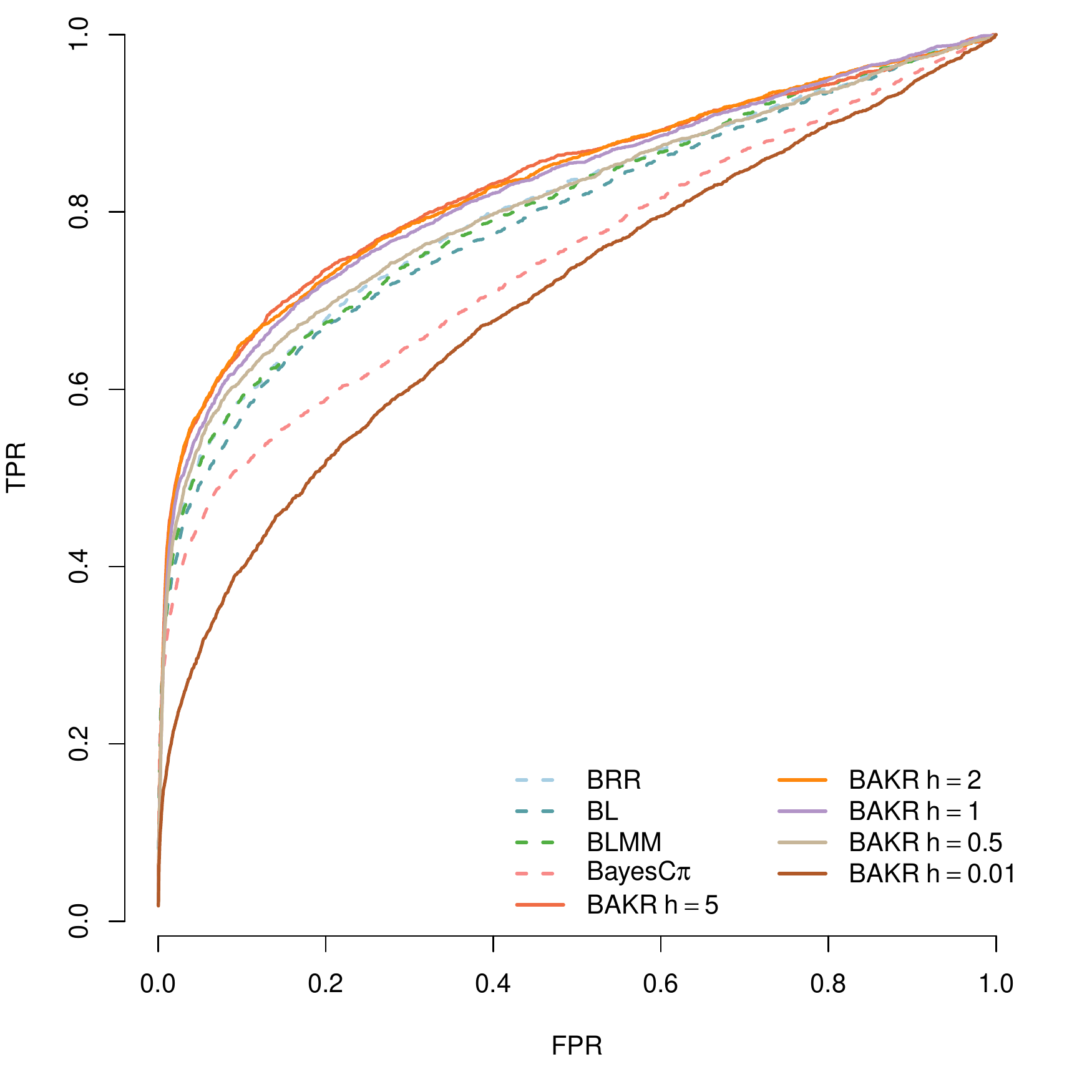}
   \label{Fig1A}
 }
 \subfigure[Scenario I ($\rho=0.2$): Group 2 SNPs]{
   \includegraphics[width = 0.4\textwidth]{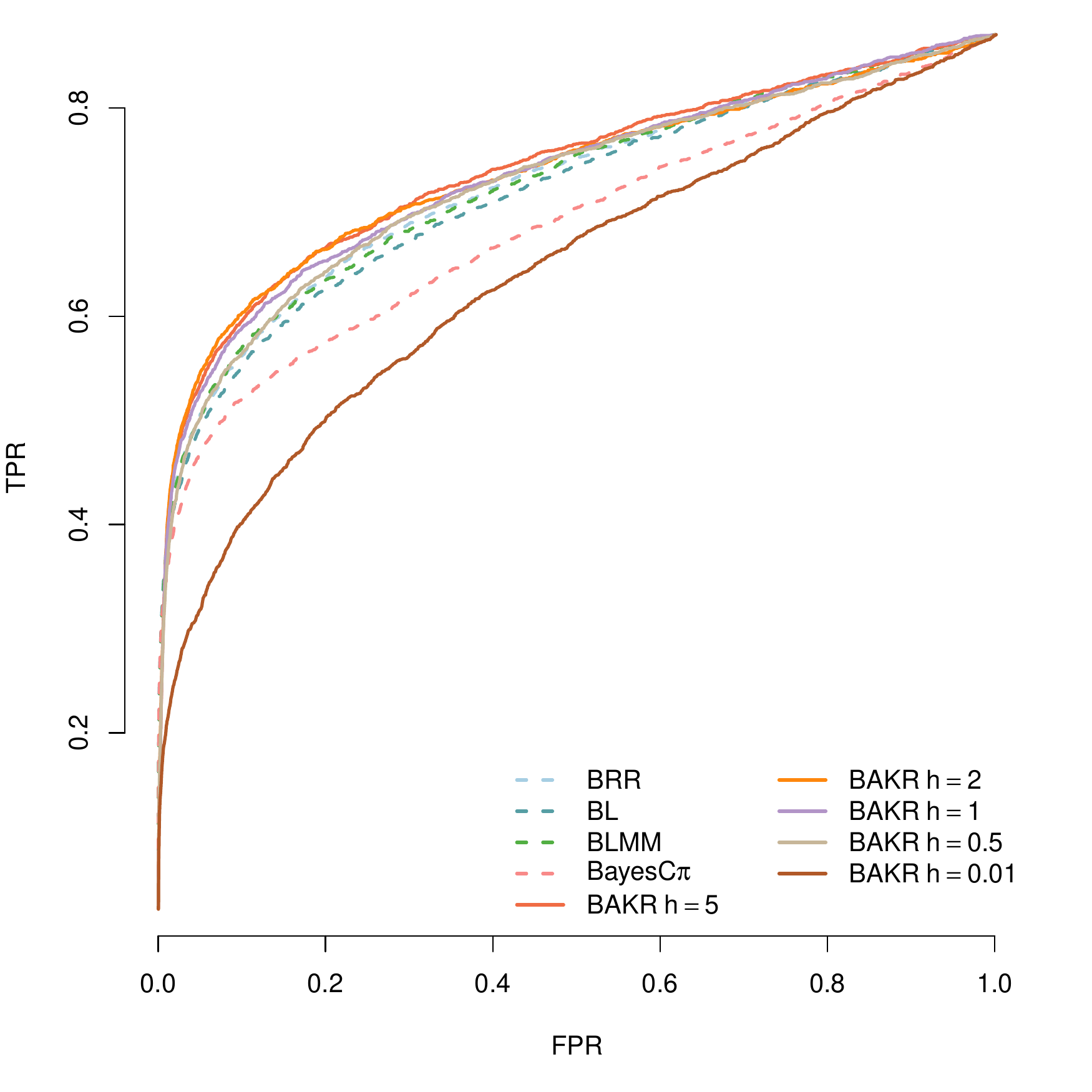}
   \label{Fig1B}
 }
 \subfigure[Scenario II ($\rho=0.8$): Group 1 SNPs]{
   \includegraphics[width = 0.4\textwidth]{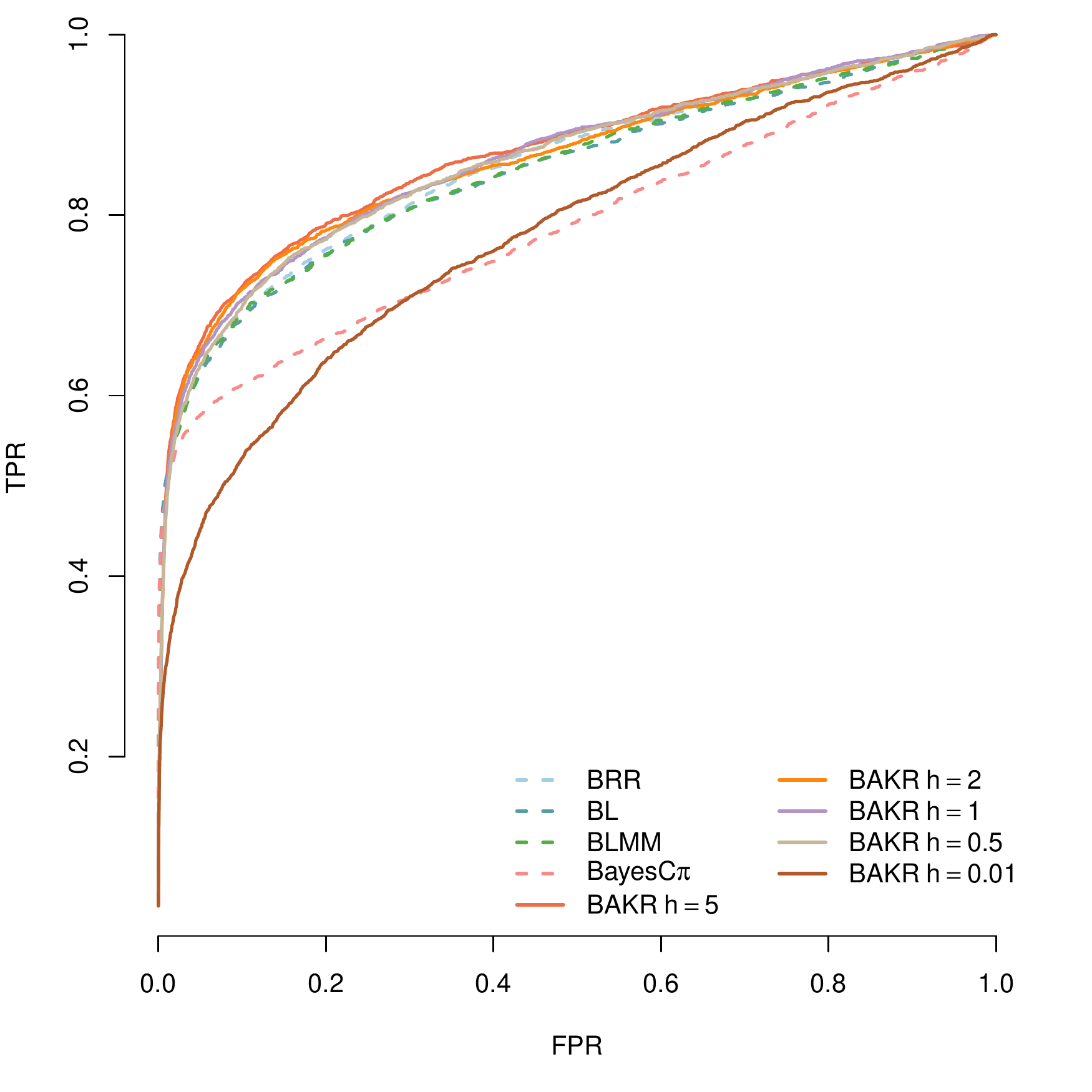}
   \label{Fig1C}
 }
 \subfigure[Scenario II ($\rho=0.8$): Group 2 SNPs]{
   \includegraphics[width = 0.4\textwidth]{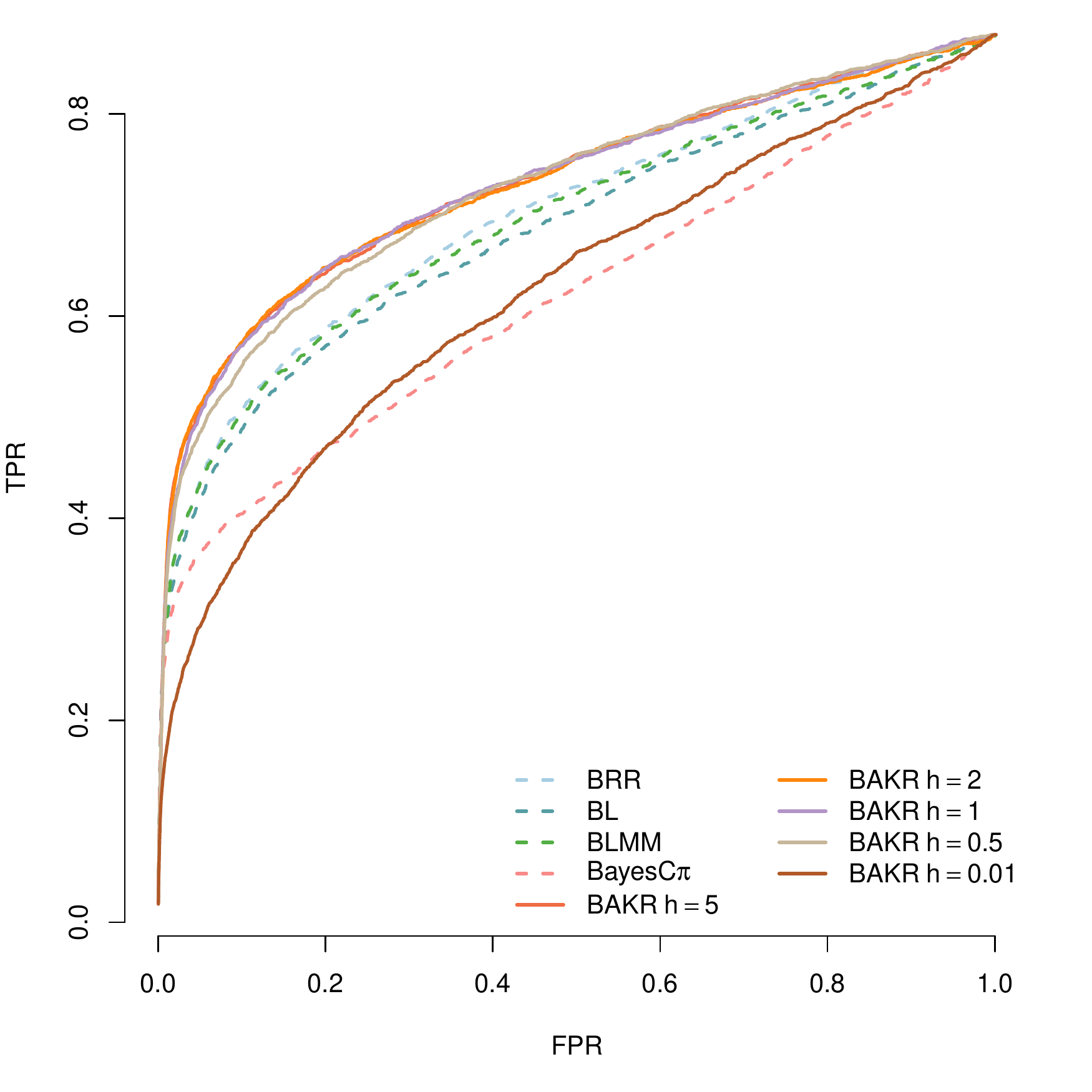}
   \label{Fig1D}
 }
\caption{Power analysis comparing Bayes Ridge (BRR), Bayes Lasso (BL), Bayes LMM (BLMM), Bayes C$\pi$, and BAKR approximating Gaussian kernels with bandwidth parameter values $h$. BAKR models are illustrated as a solid lines, while the competing models are shown as dotted lines. Group 1 SNPs are those that exhibit additive effects, while the SNPs in group 2 are those involved in interactions. Results are based on 100 different simulated datasets in each scenario.}
\label{Fig1}
\end{figure}

For each Bayesian model, we run a Gibbs sampler for 50,000 MCMC iterations with a burn-in of 25,000 posterior draws. Longer MCMC chains had little effect with respect to inference for any of these models. Figure \ref{Fig1} compares the power of all methods and their ability to detect the group 1 and 2 causal variants in both simulation scenarios. Specifically, the plots depict the true and false positive rates as the portion of causal variables discovered after prioritizing them in order according to their effect size magnitude. 

As expected, the power of the linear models to detect causal SNPs is dependent on two factors: (i) the contribution of the corresponding additive or interaction effects to the broad-sense heritability, and (ii) the underlying statistical assumptions of the methods. For example, the ability of the these methods to detect the group 1 causal additive SNPs is fairly consistent in both scenarios --- although they exhibit greater power when additive effects dominate the broad-sense heritability (e.g.~$\rho=0.8$ in scenario II). The linear models also exhibit relatively high power to detect the group 2 causal interaction SNPs when the broad-sense heritability is dominated by epistatic effects and the corresponding sizes of these effects are relatively large (e.g.~$\rho=0.2$ in scenario I). However, this power decreases when the contribution of the group 2 causal interaction SNPs is assumed to be small.   

Overall, BAKR's ability to detect causal SNPs and flexibly handle different genetic architectures is dependent on the assumed smoothness of the kernel function that it approximates. Note that our proposed method exhibits the best mapping power, in both simulation scenarios, for smoothed and ``oversmoothed'' inducing kernel bandwidths $h =$ 5, 2, 1, and 0.5. This power decays when BAKR approximates an ``undersmoothed'' kernel function containing too many spurious data artifacts (e.g.~$h=0.01$). Note that this same issue can be seen in the performance of other RKHS based models \citep{Jones:1996aa}. From an implementation perspective, there is an additional advantage in using BAKR as it utilizes a low-approximation formulation and works in $q$-dimensions with $q\le p$. Hence, it is potentially less computationally expensive than it counterparts. The other sampling-based methods we consider are forced to conduct inference in at least the original $p$-dimensional covariate space. The trade-off between computational cost and modeling performance can be seen as a function of sample size and number of covariates in Supporting Information (see Table S2).

\paragraph{Out-of-Sample Prediction.} BAKR can also be used for out-of-sample prediction. Compared with standard RKHS regression models, our proposed approach provides effect size estimates of the original covariates to facilitate prediction. Using the same two previously described simulation scenarios, we evaluate the predictive accuracy of BAKR by again comparing it with Bayes Lasso, Bayes Ridge, and Bayes LMM. Additionally, we also consider the predictive performance of a commonly used supervised RKHS learning algorithm: a support vector machine (SVM) with a Gaussian radial basis function \citep{Wahba}. We implement this model using the R package kernlab under the ``rbfdot'' model setting \citep{kernlab}. The SVM estimates its parameters deterministically and does not require MCMC.

For this simulation study, we used mean square prediction error (MSPE) to compare out-of-sample predictive accuracy. Overall numerical results are presented in Table \ref{Tab1}, and then further illustrated as boxplots in Figure \ref{Tab2} to show how methods perform while taking into account variability. Again, in Supporting Information, we conduct an empirical analysis where we assess the sensitivity of BAKR to the choice of $q$ (see Figure S2 and Table S3).

Overall, BAKR outperforms each of the models, including the SVM, in both simulation scenarios. This shows that the use of the approximate kernel matrix does not hinder BAKR's predictive accuracy in any way, even when approximating an undersmoothed kernel function (e.g.~$h=0.01$). While we cannot say with complete certainty why BAKR outperforms the SVM which uses an exact computation of the Gaussian kernel matrix, it has been shown that randomization can serve a dual role in improving both computational and statistical performance by implicitly regularizing the estimate of the kernel matrix \citep[e.g.][]{Darnell:2015aa,Rudi:2015aa}. It is also worth noting that, unlike our method, the SVM is similar to other supervised kernel techniques and cannot readily be used to infer the relevance of individual variables. Again, we want to stress that the advantage of BAKR is that it classifies as well as the best predictive methods, but also can be used to infer the relevance of variables.

\subsection{Genomic Selection in Stock Mice}

We further assess BAKR's predictive ability by analyzing 129 quantitative traits in a heterogenous stock mouse dataset \citep{Mice} from the Wellcome Trust Centre for Human Genetics (\url{http://mus.well.ox.ac.uk/mouse/HS/}). The dataset contains $n\approx2,000$ individuals and $p\approx10,000$ SNPs --- with exact numbers vary slightly depending on the phenotype. The 129 quantitative traits are classified into 6 broad categories including: behavior, diabetes, asthma, immunology, haematology, and biochemistry (see Table S4 in Supporting Information). We consider this particular dataset not only because it contains a wide variety of quantitative traits, but also because the data contains related samples. Relatedness has been shown to manifest different orders of interaction effects \citep{Hemani:2013aa}, and thus this dataset presents a realistic mix between our simulation scenarios. More specifically, each phenotypic measure represents a different disease characteristic with different levels of broad-sense heritability. Here, we compare the predictive performance of Bayes Ridge, Bayes Lasso, Bayes LMM, SVM, and BAKR approximating a Gaussian kernel with bandwidth parameter $h = 1$.

For comparison in predictive ability, we follow previous works \citep{Zhou:2013aa,Speed:2014aa} by dividing the observations into roughly equal sized training and out-of-sample test sets. We apply all five models using the genotypes only, and disregarding any other covariates. We then obtain posterior means, or point estimates, for the model effect sizes in the training data, and assess prediction performance using these estimates in the out-of-sample test set by MSPE. We perform fifty of these training and test set splits for each phenotype to obtain a robust measure of predictive performance for each method. Note that each of the 129 quantitative traits were fit and analyzed separately.

Figure \ref{Fig3} summarizes the prediction accuracy, measured by MSPE, for each of the five competing methods across the 129 quantitative mice phenotypes. We find that the parametric models consistently exhibit the worst predictive performance. More specifically, Bayes Lasso, Bayes Ridge, and Bayes LMM had average MSPEs of 1.172, 1.170, and 1.136, respectively. Consistent with our simulation study, SVM and BAKR clearly perform the best with average MSPEs of 0.911 and 0.880, respectively. In fact, SVM and BAKR were the optimal methods for all phenotypes. Also note that overall, BAKR proved to be the best model. It was the optimal method for 93\% of the phenotypes (see Table \ref{Tab2}).  

In order to better explain why BAKR (and to a lesser extent, SVM) outperformed the parametric models in each of the 129 phenotypes, we use a variance component analysis to evaluate the overall contribution of nonlinearities to the phenotypic variance explained, or PVE (see Supporting Information for details). The basic idea behind the decomposition of the PVE is through the use of a linear mixed model with multiple variance components to partition the phenotypic variance into four different categories of genetic effects: (i) an additive effects component, (ii) a pairwise interaction component, (iii) a third order interaction component, and (iv) a common environmental component which represents the effects caused by mice sharing the same cage. Disregarding any random noise, we quantify the contribution of the four genetic effects by examining the proportion of PVE (pPVE) explained by the corresponding variance component.

Figure S3 displays the PVE decomposition (see Supporting Information). In this mouse dataset, the common environmental component makes up a huge proportion of the PVE for phenotypes where cage assignment and social interaction across mice matter. In other words, the relatedness between the samples often greatly influences the variation in each trait. For example, many of the phenotypes under the ``diabetes'' and ``immunology'' categories include weight and protein measurements --- all of which depend heavily on the distribution of food and water, as well as the number of mice in each cage. Another significant finding is that additive effects are very rarely the greatest contributor in explaining variation. In particular, the pairwise and third order interaction components also contribute a large proportion to phenotypic variance. For instance, the third order interaction component explains a larger proportion of phenotypic variance than either the linear component or the pairwise component for 24 out of 129 quantitative mouse traits. 

This variance component analysis highlights the importance of accounting for population structure and modeling interaction effects greater than second order effects. The prediction results presented here are unsurprising, given that the mice are related and nonlinear effects (i.e.~effects beyond simple additivity) dominate the heritability for many of the 129 quantitative mouse phenotypes. We again stress that the advantage of BAKR lies in the explicit modeling of nonlinear relationships between covariates and the desired response, as well as in accounting for relatedness between samples. 

\subsection{Association Mapping in WTCCC Data}

We apply BAKR to an association mapping analysis of all seven diseases from the Wellcome Trust Case Control Consortium (WTCCC) 1 study \citep{WTCCC}. The WTCCC dataset has been previously used for evaluating the mapping power of other statistical methods \citep[e.g.][]{Speed:2014aa}. These data include about 14,000 cases from seven common diseases and about 3,000 shared controls, typed at a total of about 450,000 single-nucleotide polymorphisms (SNPs). The seven common diseases are bipolar disorder (BD), coronary artery disease (CAD), Crohn's disease (CD), hypertension (HT), rheumatoid arthritis (RA), type 1 diabetes (T1D), and type 2 diabetes (T2D).

For each disease, we provide a summary table which lists all significant SNPs that were detected by BAKR (see Table S5 in Supporting Information). We used the PPAA metric to assess significance of association for a particular SNP or locus (see Supporting Information for details). A twenty-fold permutation procedure was used to set a genome-wide significance threshold  by permuting the labels of the cases twenty times and choosing a trait-specific PPAA threshold that corresponds to a 5\% family-wise error rate (FWER) (see page 2 of Table S5). In Table \ref{Tab3}, we compare BAKR's findings to the loci and variants listed as ``strongly associated'' in the original WTCCC study. Table S6 displays all BAKR-discovered loci for each trait. Here, we cross-reference BAKR's results with discoveries made by other statistical methods that were implemented on the same dataset. Figure S4 displays manhattan plots of our genome-wide scan for all seven diseases (again see Supporting Information).

Overall, BAKR identified 29 significantly associated genomic regions --- 14 of which were highlighted in the original WTCCC study as having strong associations, and 3 others that were highlighted in other studies which analyzed the same dataset. BAKR missed 6 genomic regions that were identified as strongly associated in the original WTCCC study, but was able to discover 12 new loci in five of the seven diseases: CD, HT, RA, T1D, and T2D. Many of these findings also have potentially functional relevance in the context of the given diseases. For example, in CD, variants spanning from 70.20Mb-70.29Mb on chromosome 10 were detected by BAKR as being associated with Crohn's disease. The leading significant SNP in this region with the highest PPAA is rs2579176. This variant, in particular, has been reported as being upstream of \textit{DLG5}, a gene which has been found to be associated with perianal Crohn's disease \citep{Ridder:2007aa}. This gene was also validated as being a member of a pairwise genetic interaction that is very influential in the cause of the trait and hard to detect \citep{Zhang:2012aa}. Complete details of all potentially novel loci discovered by BAKR can be found in the Supporting Information. 


\section{Discussion}\label{sec5}


In the present study, we used an approximate Gaussian kernel function in the BAKR model. The Gaussian kernel includes all all higher-order interaction components, where the contribution of the terms decays polynomially with the order of interaction. BAKR can be applied to any shift-invariant kernel. Hence, in the context of genetics, it is of particular interest to match the shift-invariant kernel function with the magnitude of the higher-order interaction(s) that best capitulate the underlying genetic architecture for a given trait or phenotype. For spatial applications, kernel functions such as the Mat\'{e}rn covariance function would be of interest. BAKR can also be extended to model multiple responses, and an interesting future direction would be to take advantage of correlations in the multiple response variables to increase power and identify pleiotropic interaction effects.

The greatest limitation of BAKR is that while it provides an analog of effect size for nonlinear functions, it cannot be used to directly identify the component that drives individual variable associations. In particular, after identifying an associated variable, it is often unclear which component (i.e.~linear vs.~nonlinear) drives the association. Thus, despite being able to identify predictors that are associated to a response in a nonlinear fashion, BAKR is unable to directly identify the detailed interaction effects. There are two possible approaches available within the BAKR framework to address this concern. One approach is to use the posterior samples of the effect size analogs to infer covariance and partial correlation structure. This partial correlation structure can then be used to posit putative interactions. Another approach is to use BAKR as a screen to select candidate variables and then test for pairwise interactions between the variables that are most marginally significant. There are also computational constraints and while BAKR can easily deal with tens of thousands of samples and millions of variables, scaling BAKR to hundreds of thousands or millions of samples will require algorithmic innovations.

\section*{Software Availability}

Software for implementing Bayesian approximate kernel regression (BAKR) is carried out in R and Rcpp code, which is freely available at \url{https://github.com/lorinanthony/BAKR}.


\section*{Supporting Information}

Supporting Information is available for download at \url{https://github.com/lorinanthony/BAKR/tree/master/SI}.


\section*{Acknowledgements}
LC, KCW, XZ, and SM would like to thank Mike West, Elizabeth R. Hauser, and Jenny Tung for useful conversations and suggestions. LC is supported by the National Science Foundation Graduate Research Program under Grant No.~DGF-1106401. KCW would like to acknowledge the support of the NIH BIRCWH Program, a V Scholar Award from the V Foundation for Cancer Research, a Liz Tilberis Early Career Award from the Ovarian Cancer Research Fund, a Lloyd Trust Translational Research Award, and a Stewart Trust Fellowship. XZ would like to acknowledge the support of NIH Grants R01HG009124, R01HL117626 (PI Abecasis), R21ES024834 (PI Pierce), R01HL133221 (PI Smith), and a grant from the Foundation for the National Institutes of Health through the Accelerating Medicines Partnership (BOEH15AMP, co-PIs Boehnke and Abecasis). SM would like to acknowledge the support of grants NSF IIS-1546331, NSF DMS-1418261, NSF IIS-1320357, NSF DMS-1045153, and NSF DMS-1613261. Any opinions, findings, and conclusions or recommendations expressed in this material are those of the author(s) and do not necessarily reflect the views of any of the funders. This study makes use of data generated by the Wellcome Trust Case Control Consortium (WTCCC). A full list of the investigators who contributed to the generation of the data is available from \url{www.wtccc.org.uk}. Funding for the WTCCC project was provided by the Wellcome Trust under award 076113 and 085475.


\section*{Figures and Tables}

\begin{table}[H]
\centering
\begin{tabular}{c|c|cccc|cccccc}
\hline
& & \multicolumn{4}{c}{Other Methods} & \multicolumn{5}{|c}{BAKR Models}\\
\hline
 & Scenario & BRR & BL & BLMM & SVM & $h=5$ & $h=2$ & $h=1$ & $h=0.5$ & $h=0.01$ \\
\hline
\multirow{4}{*}{\shortstack{MSPE\\ \\(SD)}} & \multirow{2}{*}{I} & 1.806 & 1.683 & 1.609 & 0.876 & \textbf{0.722} & \textbf{0.722} & 0.724 & 0.725 & 0.844\\
& & (1.45) & (1.84) & (1.32) & (0.18) & \textbf{(0.21)} & \textbf{(0.21)} & (0.21) & (0.21) & (0.17)\\\cline{2-11}
& \multirow{2}{*}{II} & 1.825 & 1.514 & 1.429 & 0.895 & 0.721 & \textbf{0.719} & \textbf{0.719} & 0.724 & 0.853\\
& & (1.76) & (1.29) & (1.22) & (0.16) & (0.19) & \textbf{(0.17)} & \textbf{(0.17)} & (0.17) & (0.14)\\
\hline
\end{tabular}
\caption{Comparison of the mean square prediction error (MSPEs) for Bayes Ridge (BRR), Bayes Lasso (BL), Bayes LMM (BLMM), Support Vector Machine (SVM), and BAKR approximating Gaussian kernels with bandwidth parameter values $h$. Values in bold represent the method with the lowest average MSPE. Standard errors across these replicates for each model are given the parentheses.}
\label{Tab1}
\end{table}

\begin{figure}[H]
\center{
\includegraphics[width = \textwidth]{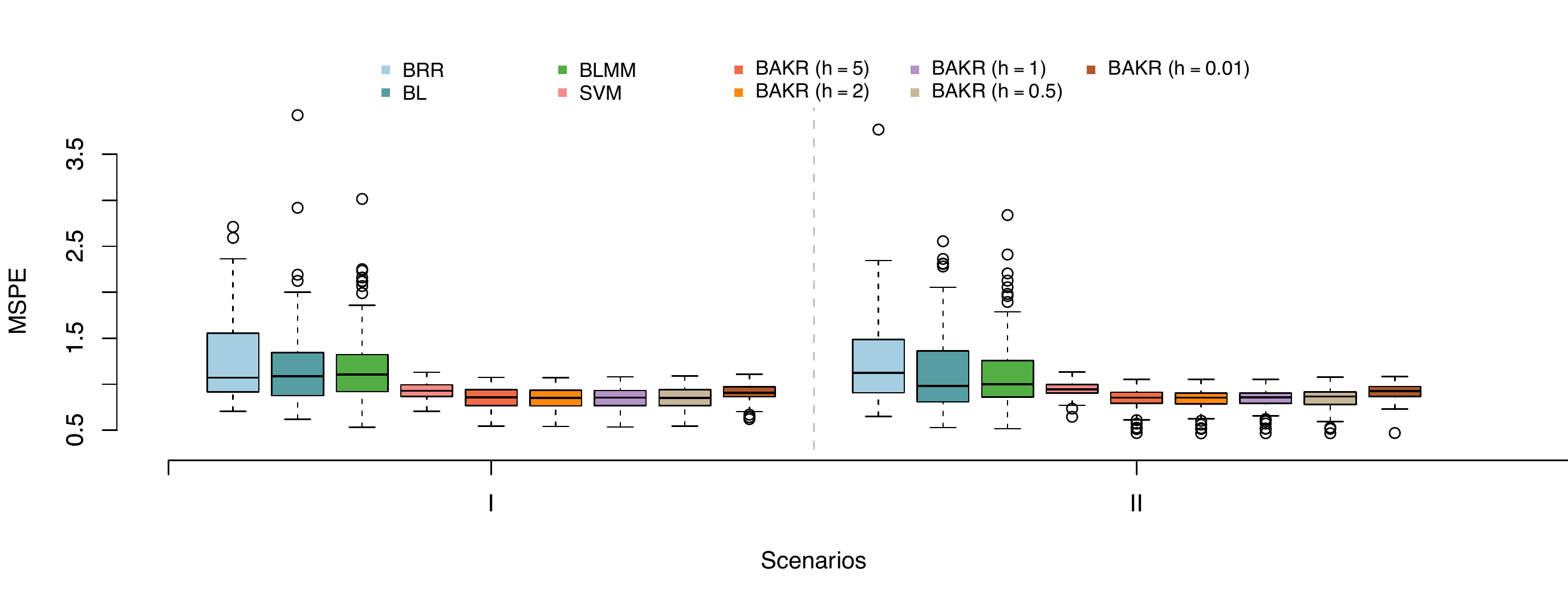}
}
\caption{Comparison of the mean square prediction error (MSPEs) for Bayes Ridge (BRR), Bayes Lasso (BL), Bayes LMM (BLMM), Support Vector Machine (SVM), and BAKR approximating Gaussian kernels with bandwidth parameter values $h=\{5,2,1,0.5,0.01\}$. In Scenario I, pairwise interactions make up 80\% of the broad-sense heritability (i.e.~$\rho=0.2$). In Scenario II, additive effects dominate 80\% of the broad-sense heritability (i.e.~$\rho=0.8$). These results are based on 100 different simulated datasets in both scenarios.}
\label{Fig2}
\end{figure}

\begin{table}[H]
\centering
\begin{tabular}{c|ccccc}
\hline
 & BRR & BL & BLMM & SVM & BAKR ($h=1$) \\
\hline
MSPE (SD) & 1.172 (0.18) & 1.170 (0.19) & 1.136 (0.18) & 0.911 (0.13) & \textbf{0.880 (0.10)}\\
Optimal\% & 0 & 0 & 0 & 0.07 & \textbf{0.93}\\
\hline
\end{tabular}
\caption{Comparison of overall mean square prediction errors (MSPEs) for each of the five considered models across of all 129 quantitative stock mice phenotypes. The proportion of phenotypes for which a method exhibits the lowest MSPE is denoted as Optimal\%. Values in bold represent the approach with the best (and most robust) performance. Standard errors for each model are given the parentheses.}
\label{Tab2}
\end{table}

\begin{figure}[H]
\center{
\includegraphics[width = \textwidth]{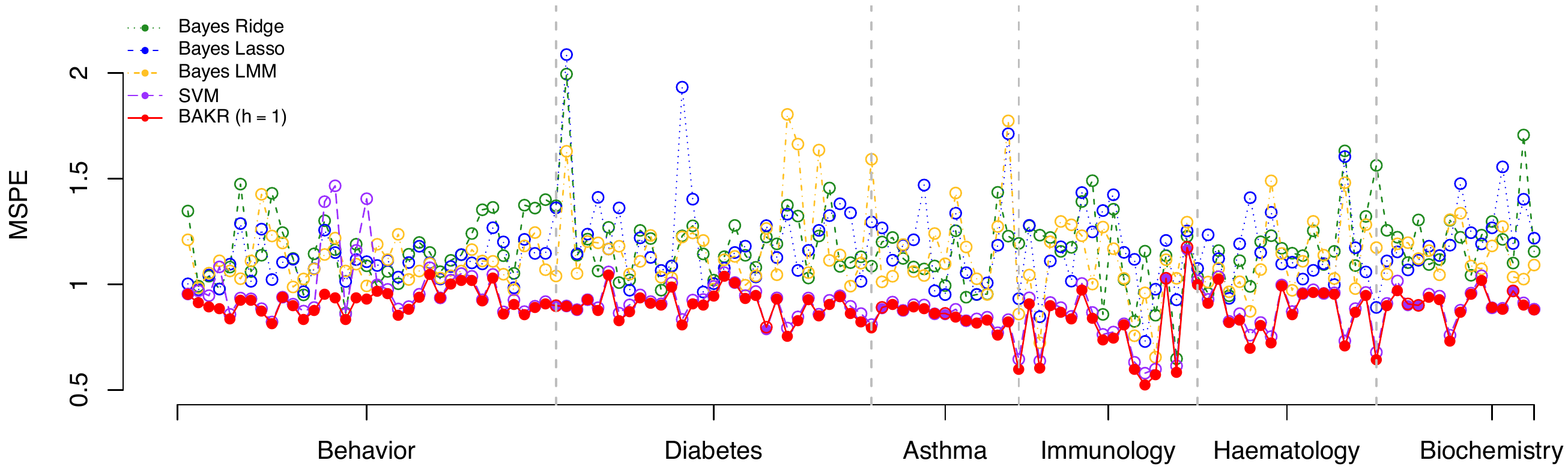}
}
\caption{Comparison of mean square prediction error (MSPE) for Bayes Ridge (BRR), Bayes Lasso (BL), Bayes LMM (BLMM), Support Vector Machine (SVM), and BAKR approximating a Gaussian kernel with bandwidth parameter value $h=1$. These results are based on fifty fold cross-validation, with the most accurate method for each phenotype is marked by a solid circle.}
\label{Fig3}
\end{figure}

\begin{table}[H]
\centering
\begin{tabular}{ccccc}
  \hline
Disease & \# Sig.~SNPs & \# Sig.~Regions & \# Add.~Regions & \# Missed Regions\\
  \hline
BD & 0 & 0 & 0 & 1\\
CAD & 15 & 1 & 0 & 0\\
CD & 80 & 8 & 1 & 2\\ 
HT & 20 & 1 & 0 & 0\\ 
RA & 177 & 4 & 1 & 0\\ 
T1D & 440 & 12 & 7 & 1\\
T2D & 23 & 3 & 2 & 2\\ 
\hline
\end{tabular}
\vspace{1 em}
\caption{A summary table comparing the SNPs and loci discovered by BAKR with the findings reported as showing evidence of association in the original WTCCC study. Significance of SNPs was determined using the PPAA metric and a 5\% FWER threshold.}
\label{Tab3}
\end{table}




\clearpage
\newpage
\bibliographystyle{chicago}  
\bibliography{BAKR_Ref} 


\end{document}